\documentclass[prb,showpacs,preprintnumbers,amsmath,amssymb,letterpaper,twocolumn]{revtex4}

\usepackage{graphicx}
\usepackage{dcolumn}
\usepackage{bm}
\usepackage{epstopdf}
\usepackage{latexsym}
\usepackage{epsfig}
\usepackage{verbatim}
\usepackage{hyperref}
\usepackage{color}

\begin{document}

\title{Asymmetric Avalanches in the Condensate of a Zeeman-limited Superconductor}

\author{J. C. Prestigiacomo, T.J. Liu, and P.W. Adams}
\affiliation{Department of Physics and Astronomy, Louisiana State University, Baton Rouge, Louisiana 70803, USA}

\date{\today}

\begin{abstract}
 We report the non-equilibrium behavior of disordered superconducting Al films in high Zeeman fields.   We have measured the tunneling density of states of the films through the first-order Zeeman critical field transition.  We find that films with sheet resistances of a few hundred ohms exhibit large avalanche-like collapses of the condensate on the superheating branch of the critical field hysteresis loop.  In contrast, the transition back into the superconducting phase (i.e., along the supercooling branch) is always continuous. The fact that the condensate follows an unstable trajectory to the normal state suggests that the order parameter in the hysteretic regime is not homogeneous.
 \end{abstract}


\maketitle

 \section{Introduction}
 
 Spin-imbalanced superconductivity is an historically important problem that remains at the forefront of condensed matter physics \cite{Quay}.  By the late 1960's it was known that a Zeeman field could induce a spatially modulated order parameter in a spin singlet superconductor, i.e. the Ferrel-Fulde-Larkin-Ovchinnikov (FFLO) state \cite{FF,LO}.   Unfortunately, the observation of the FFLO phase has historically been hampered by its exquisite sensitivity to disorder.  But, nevertheless, over the last decade compelling thermodynamic evidence for its existence has emerged from studies of ultra-low impurity bulk superconductors such as the heavy fermion inter-metallic CeCoIn$_5$ \cite{Radovan,Kout} and the layered  organic superconductors \cite{Beyer, Coniglio, Bergk}.  A cold atomic gas analog of FFLO has also been proposed \cite{Sheehy,Liao}.  In this Letter, we present tunneling density of states evidence for a inhomogeneous superconducting phase in the hysteretic region of the Zeeman critical field transition in disordered ultra-thin Al films.   We observe asymmetric avalanche behavior in the condensate that we believe is a manifestation of a disordered {\it remnant} of FFLO correlations \cite{Yang}.  The avalanches arise from the convolution of low dimensionality, disorder, Zeeman-splitting, and spin-singlet pairing.

	In this study a magnetic field was applied parallel to the surface of superconducting Al films, having thicknesses that were approximately 5 times smaller than the coherence length ($\xi\sim13$ nm).  In this limit, the orbital response to the field is suppressed, and the transition occurs when the Zeeman splitting is of the order of the superconducting gap $\Delta_0$ \cite{Fulde}.  The conventional picture is that this Zeeman mediated transition, which is often referred to as the spin-paramagnetic (S-P) transition, occurs between a BCS ground state with a homogenous order parameter and a polarized Fermi liquid normal state. At low temperatures the Zeeman critical field is expected to be near the Clogston-Chandrasekhar \cite{Clogston,Chandra} value $H_c=\Delta_0/\sqrt{2}\mu_{\rm B}$.
	
\section{Sample Preparation}	

	Samples were fabricated by first preparing aluminum films from 99.999\% Al targets via e-beam deposition onto fire polished glass substrates held at 84 K.  The deposition rate was held constant at 1 {\AA}/s in a 0.1 $\mu$Torr vacuum.  Films with thicknesses ranging from $t=20\to30$ {\AA} had normal-state sheet resistances that ranged from  $R_n=5.5$ k$\Omega$/sq to 80 $\Omega$/sq at 80 mK, respectively, and a disorder-independent superconducting transition temperature of $T_c\sim2.7$ K.   Warming the films to 295 K after deposition and then exposing them to ambient conditions for 10-20 min formed a thin native oxide, which served as the tunneling barrier.  A 90-{\AA}-thick Al counterelectrode (CE) deposited on top of the oxide created a junction area of about 1 x 1 mm$^2$.  Due to finite thickness effects, the CE parallel critical field was near 3 T as compared to the 6 T critical field of the films.  In the data present below the applied field was well above 3 T and the CE was in the normal state.  Thus, all of the tunneling spectra are of the superconductor-insulator-normal type.  The barrier resistances ranged from 1 k$\Omega$ to 10 k$\Omega$ depending on the thickness of the electrode, exposure time, and other factors.  Only junctions with barrier resistances much higher than the films' resistance were used.  Transport and tunneling data were collected via a 4-probe configuration with a lock-in amplifier.  The films were cooled using a dilution refrigerator equipped with a mechanical rotator allowing us to align the films to within $0.1^\circ$ of parallel field.

\section{Experimental Results}

Previous transport measurements of the parallel critical field behavior of Al films similar to the ones used in this study revealed a hysteretic first-order transition at temperatures below 500 mK \cite{Wu1,Wu2}. Discrete jumps in resistance were observed at the edges of the hysteresis loops of films \cite{Wu2}.  Since the films in these earlier studies had thicknesses much less than the coherence length, the jumps were interpreted as non-flux avalanches.  However, it was unclear whether or not the observed avalanches actually represented the behavior of the condensate.  For instance, a sample will have zero resistance so long as there is at least one superconducting filamentary path along its length. Therefore, avalanches in resistance do not necessarily correspond to avalanche-like changes in the order parameter.

We have employed tunneling density of states (DOS) to probe the non-equilibrium behavior in the hysteretic region of the S-P transition.  At low temperatures the tunneling conductance is proportional to the density of electronic states (DOS) of the film \cite{Tinkham}. Since planar tunneling is an areal microscopic probe of the condensate, it is relatively insensitive to filamentary superconductivity.  Therefore, tunneling offers the opportunity to determine the ultimate origin of the avalanche events.

\begin{figure}
\begin{flushleft}
\includegraphics[width=.44\textwidth]{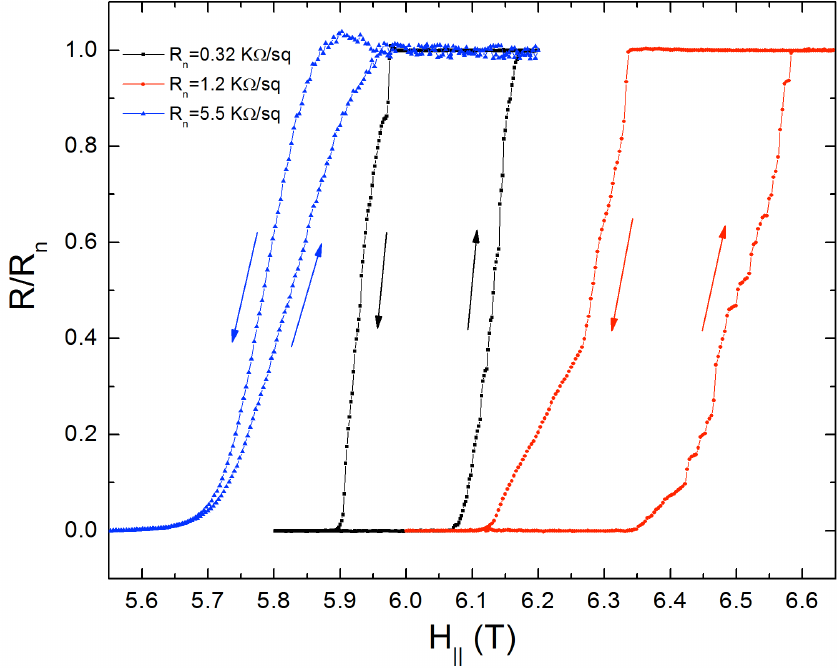}\end{flushleft}
\caption{\label{Fig1} Parallel critical field transitions measured at 60 mK for three Al films of varying normal state sheet resistance.  The sheet resistance of each film has been normalized by its normal state value.}
\end{figure}

Figure\ \ref{Fig1} shows the resistive parallel critical field transition of three Al films of varying normal state resistance.  In all three of the films the transition is first-order, but the detailed character of the hysteresis loops is clearly a function of the sheet resistance.  The lowest resistance film exhibits the highest critical field and the widest hysteresis loop.   It also shows the strongest propensity for avalanches.  Indeed,  the upsweep branch (superheating branch) of the $R_n=0.32$ k$\Omega$/sq film has many discontinuous jumps in resistance and is generally more ragged than the corresponding branch of the $R_n=5.5$ k$\Omega$/sq film.  These data suggest that the avalanche behavior is limited to films with modest disorder and that it is almost completely suppressed once the film resistance is of the order of the superconducting quantum resistance $\hbar/(4e^2)\sim6.5$ k$\Omega$/sq \cite{Baturina,Hollen}.  For this reason we have focused our tunneling studies on moderately disordered films having normal state sheet resistances of a few hundred ohms.  In this resistance range the coherence length is $\xi\sim13$ nm \cite{Butko} and the electron mean-free-path is $l\sim$1 - 2 nm \cite{Wu}.

\begin{figure}
\begin{flushleft}
\includegraphics[width=.44\textwidth]{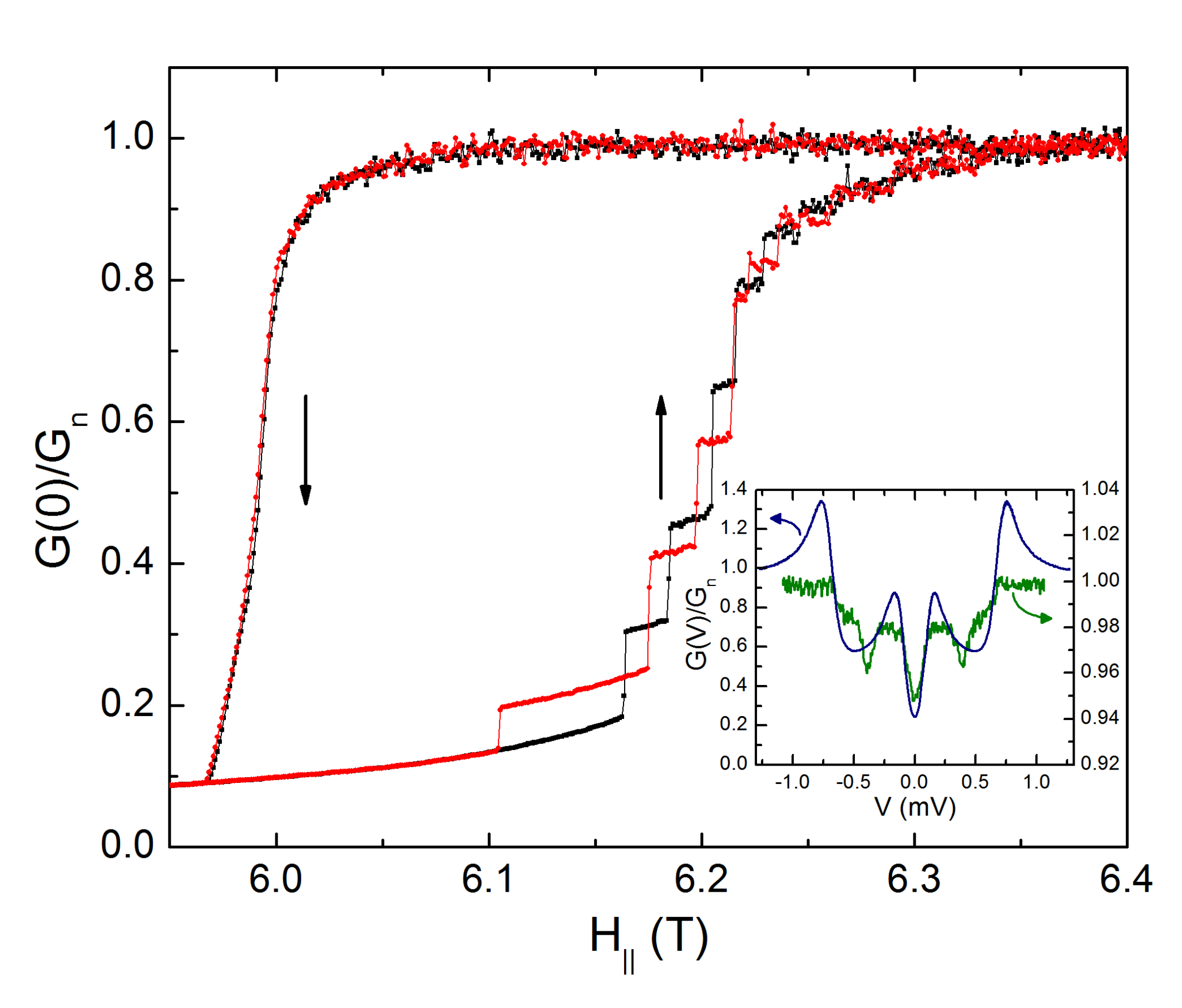}\end{flushleft}
\caption{\label{Fig2} The zero bias tunneling conductance at $T=52$ mK.  The data is normalize by the zero-bias normal state conductance, G$_n$.  The spectrum was obtained from a $R_n=540~\Omega$/sq Al film using a non-superconducting Al counter-electrode, with the magnetic field applied parallel to the film surface.  The red and black lines represent two separate sweeps through the hysteresis loop.  The arrows depict the field sweep direction.  Inset: The tunnel density of states spectrum of a 200 $\Omega$/sq film in a subcritical (5.4 T, blue) and a supercritical (5.9 T, green) parallel magnetic field at 80 mK.}
\end{figure}

The inset of Fig.\ \ref{Fig2} displays tunneling DOS spectra of a 200 $\Omega$/sq  Al film in a subcritical and supercritical parallel field, respectively. The Zeeman splitting of the usual BCS DOS superconducting peaks is clearly visible in the superconducting spectrum, with spin-up and spin-down coherence peaks positioned on either side of the Fermi energy \cite{Fulde} at  $eV_0=\Delta_0\pm\mu_{\rm B}H_\parallel$ , where $e$ is the electron charge and $\Delta_0/e=0.46$ mV.    Above the transition, the superconducting gap closes and is replaced by the two-dimensional $\ln(V)$ zero-bias anomaly \cite{Altshuler}.  Also present in the normal-state spectrum are satellite anomalies, seen as small dips positioned on either side of $V=0$.   These are manifestations of the pairing resonance \cite{Catelani}.  

In order to probe the condensate behavior in the transition region, we have measured the zero-bias tunneling conductance as a function of applied field.  This gives us a direct probe of the quasiparticle DOS at the Fermi energy in the transition region. In the main panel of Fig.\ \ref{Fig2} we plot the zero-bias tunneling conductance of a  540 $\Omega$/sq film as a function of parallel field.  These data span the S-P transition and were obtained by making two identical high resolution hysteresis traces at a magnetic field sweep rate of 20 G/s. The hysteresis width in Fig.\ \ref{Fig2} is comparable to what we observed in transport, but the avalanche behavior is somewhat different. Note that there are clear step-like features on the upsweep trace (superheating branch) but none on the down-sweep (super-cooling branch).  We believe that these steps are conclusive evidence that the avalanches occur in the condensate and that they involve superconducting regions that have lateral dimensions much greater than the superconducting coherence length.  Of course, it is possible that there may also be much smaller avalanches that cannot be resolved by our tunneling probe. 

The asymmetric avalanche behavior was seen in all of the moderately disordered samples we measured.  We believe that avalanches were missed in previous tunneling density of states studies \cite{Butko2} for two reasons.  First, the field sweep rate must be sufficiently slow so to allow the system to relax to an avalanche event.  Second, the phase sensitive detection must have enough bandwidth to resolve the jumps in the density of states.  If one is not specifically looking for avalanches, then it is easy to dismiss them as sporadic noise.

The asymmetry of avalanche behavior is unusual. For instance, in the Barkhausen effect non-thermal magnetic domain wall jumps produce avalanche-like features in the magnetization loop of ferromagnetic alloys \cite{Barkhausen1,Barkhausen2}.  However, Barkhausen avalanches are distributed symmetrically across both branches of the hysteresis loop.  Similarly, thermally induced martensitic transitions exhibit avalanches when the sample is either cooled or heated through the critical region \cite{martensitic}. The data in Fig.\ \ref{Fig2} suggest that, when under the influence of a pure Zeeman field, the system cannot find a continuous path out of the superconducting phase, but can make a continuous transition from the normal state to the superconducting phase.  Interestingly, the avalanches on the superheating branch can be completely suppressed by tilting the film out of parallel orientation by as little as $1.5^\circ$, although the hysteresis remains mostly intact at such a small misalignment.   A misalignment of $1.5^\circ$ corresponds to a perpendicular magnetic field of $H_\perp\sim0.15$ T, which is an order of magnitude smaller than $H_{c2}$.

A systematic study of the influence of disorder on the details of the avalanche behavior has not been done, but we have made tunneling measurements on films with normal state resistances that are near the threshold of the superconductor-insulator transition \cite{SI} ({\it i.e.}, $R_n\sim h/4e^2\approx 6.5$ k$\Omega$/sq).  These highly disordered films have rather broad critical field transitions (see Fig.\ \ref{Fig1}) and a finite tri-critical point, but they do not exhibit avalanches.   

\begin{figure}
\begin{flushleft}
\includegraphics[width=.44\textwidth]{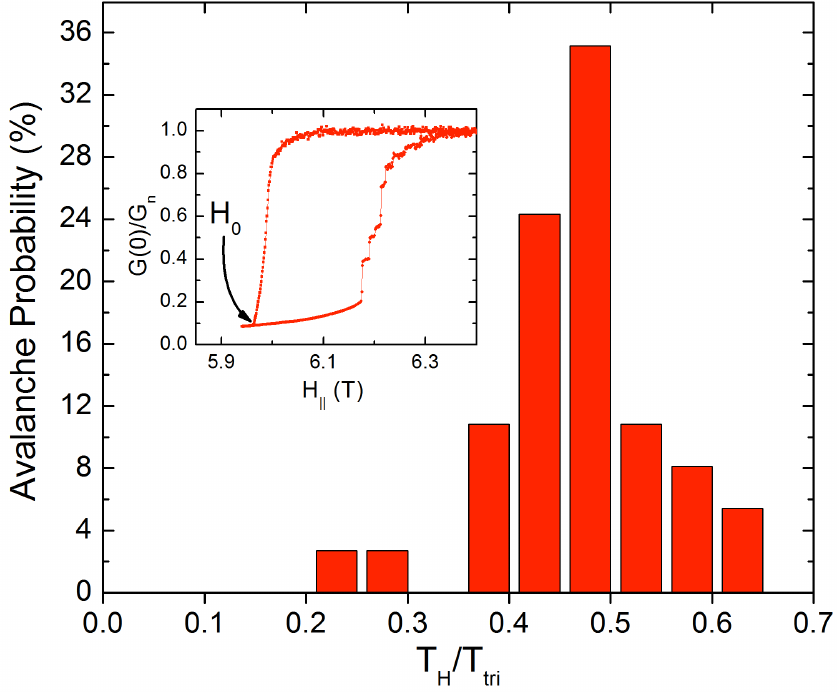}\end{flushleft}
\caption{\label{Fig3} A histogram of the avalanche probability as a function of the normalized non-equilibrium Zeeman temperature, see text. The avalanche statistics were collected from six tunneling DOS hysteresis loops that were measured consecutively with identical field sweep protocols.  Inset: A typical hysteresis loop used to collect the statistics in the main figure.  The arrow indicates the low-field closure point of the loop, $H_0$.}
\end{figure}

\begin{figure}
\begin{flushleft}
\includegraphics[width=.44\textwidth]{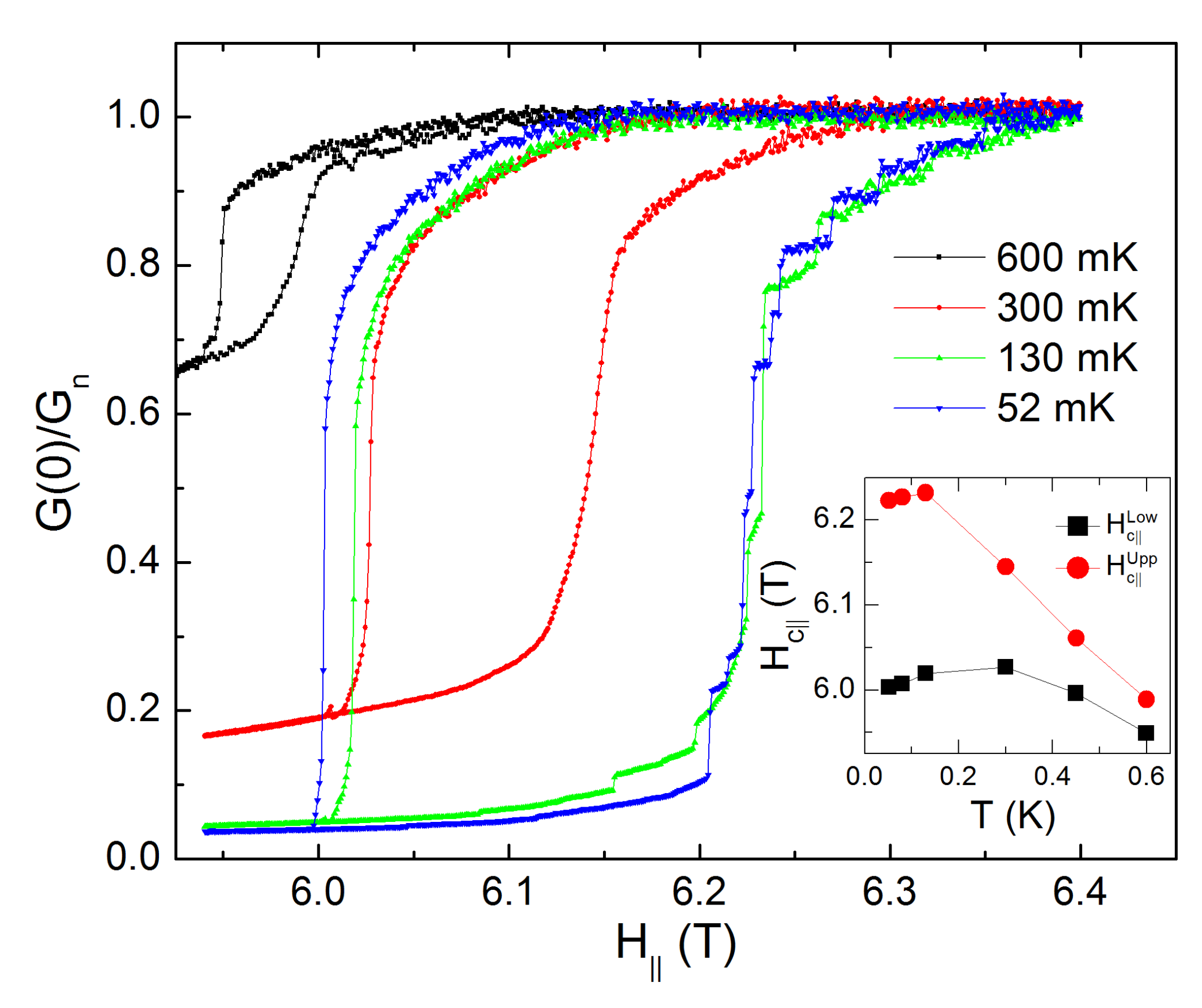}\end{flushleft}
\caption{\label{Fig4}  Hysteresis in the normalized zero-bias tunneling conductance of the film in Fig.\ \ref{Fig2} at various $T<T_{\rm Tri}$.  Inset:  The upper and lower parallel critical fields as a function of temperature determined from the midpoint of the critical field transitions as obtained from the zero-bias tunneling conductance.}
\end{figure}

In Fig.\ \ref{Fig3}, we show a histogram of the probability of an avalanche event versus the non-equilibrium Zeeman energy. The data were compiled from six consecutive cycles through the hysteresis loop.   We define an effective non-equilibrium Zeeman temperature as $T_H=2\mu_{\rm B}\Delta H/k_{\rm B}$, where $\Delta H=H-H_0$ is the extent to which the field has been ramped past the low-field closure point of hysteresis loop $H_0$, see inset of Fig.\ \ref{Fig3}.  We believe that, as the field is ramped up into the hysteretic region, the condensate is pushed further and further out of equilibrium by the increasing Zeeman splitting.  The parameter $T_H$ reflects this ``stress'', which ultimately leads to an avalanche.  In producing the histogram, we normalized $T_H$ by the tricritical point temperature $T_{\rm Tri}=730$ mK.   Note that the avalanche probability distribution of Fig.\ \ref{Fig3} peaks near  $T_H/T_{\rm Tri}\sim 0.45$, which corresponds to $T_H^{peak}=330$ mK.  Interestingly, this value of $T_H^{peak}$ is consistent with our observation that the avalanches disappear at temperatures above $\sim300$ mK, as can be seen Fig.\ \ref{Fig4}.   

\begin{figure}
\begin{flushleft}
\includegraphics[width=.44\textwidth]{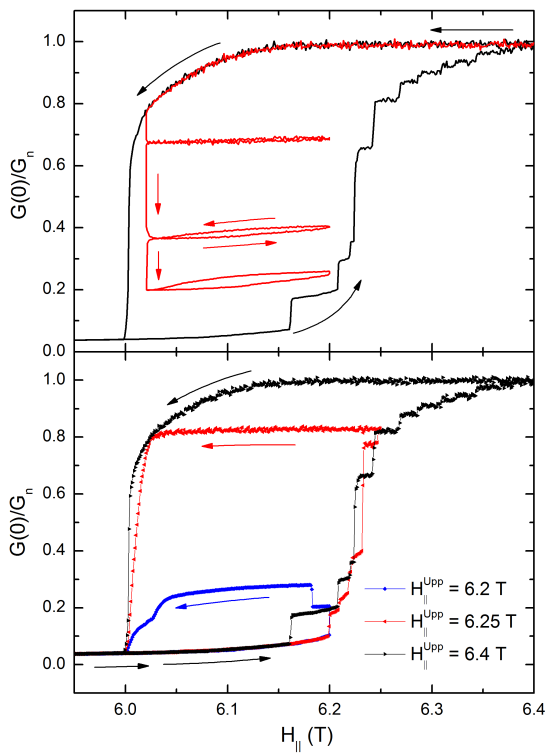}\end{flushleft}
\caption{\label{Fig5}  Minor hysteresis loops of the zero-bias tunneling conductance for the film of Fig.\ \ref{Fig2}.  The arrows indicate the field sweep direction.  Upper panel:  Minor loops off of the super-cooling branch taken in sequence with a 30 s interval between each. The upper minor loop was obtained by first halting the initial down-sweep at 6.02 T for 30 s.  The field was then ramped up to 6.2 T and back to 6.02 T.  This process was repeated twice more, which resulted in the lower two minor loops.    Lower panel:  Minor loops off of the superheating branch.  With each subsequent loop, the field was swept closer to the upper critical field and then returned to the initial subcritical field of 5.9 T.  The minor loop traces are labeled by their respective maximum field.}
\end{figure}

\section{Discussion}

Although, it is generally accepted that even modest levels of disorder destroys the FFLO phase \cite{Radovan}, recent Hubbard calculations suggest that some vestiges of the FFLO state remain at finite impurity densities \cite{Loh,Yang}.  Because there is no long range FFLO order, only local modulations of the pairing amplitude persist.  It is unclear what effect these local modulations have on the details of the phase diagram \cite{Yang}, see inset of Fig.\ \ref{Fig4}.  Nevertheless, we believe that, in the critical field region, the order parameter develops positive and negative regions that are separated by domain walls containing Andreev bound states.  These domain walls conform to the local disorder landscape so as to minimize the free energy of the system.  The tunneling data show that the system can readily optimize its domain wall configuration when transitioning from the normal state to the superconducting state, but once the configuration is formed, the domain walls remain pinned over a finite range of Zeeman fields.  Consequently, as one approaches the superheating critical field branch, avalanches occur as a result of a conflagration of domain wall de-pinning events.

The asymmetry of the non-equilibrium behavior is also evident in the minor hysteresis loops shown in Fig.\ \ref{Fig5}.  The upper panel shows a series of minor loops that were initiated from the super-cooling branch.  Each loop was swept out in succession at 30 s intervals.   The precipitous drop in tunnel conductance that precedes each of the minor loops is due to temporal relaxation on the supercooling branch.  Note that, once off the branch, the minor loops exhibit very little relaxation, and all three loops return to their starting point and, hence, exhibit return point memory \cite{Sethna}.   In contrast, the minor loops initiated from the superheating branch, shown in the lower panel of Fig.\ \ref{Fig5}, are interspersed with multiple avalanches.  Consequently, these do not display return point memory.

In summary, we have demonstrated that the condensate of a moderately disordered low-spin orbit scattering, BCS  superconductor exhibits asymmetric avalanche behavior near the Zeeman critical field. The avalanches represent irreversible collapses of macroscopic regions of superconductivity, but are not associated with magnetic flux jumps.   Future studies of other low atomic mass superconductors, which would presumably have differing film morphologies and superconducting parameters, should prove invaluable in further elucidating the origins and characteristics of the non-equilibrium behavior of the Zeeman-limited superconducting state.


We thank Nandini Trivedi and Gianluigi Catelani for their suggestions and insights.  This work was supported by the U.S. Department of Energy, Office of Science, Basic Energy Sciences, under Award No.\ DE-FG02-07ER46420.

\end{document}